# Revealing the Compositional Control of Electrical, Mechanical, Optical, and Physical Properties of Inorganic Glasses

*R. Ravinder[1], Suresh Bishnoi[1], Mohd Zaki[1], N. M. Anoop Krishnan[1,2]\**

[1]Department of Civil Engineering, Indian Institute of Technology Delhi, Hauz Khas, New Delhi 110016, India

[2]School of Artificial Intelligence, Indian Institute of Technology Delhi, Hauz Khas, New Delhi 110016, India

*Corresponding author: N. M. A. Krishnan ([krishnan@iitd.ac.in](krishnan@iitd.ac.in))*

## Abstract

Inorganic glasses, produced by the melt-quenching of a concoction of minerals, compounds, and elements, can possess unique optical and elastic properties along with excellent chemical, and thermal durability. Despite the ubiquitous use of glasses for critical applications such as touchscreen panels, windshields, bioactive implants, optical fibers and sensors, kitchen and laboratory glassware, thermal insulators, nuclear waste immobilization, optical lenses, and solid electrolytes, their composition–structure–property relationships remain poorly understood. Here, exploiting largescale experimental data on inorganic glasses and explainable machine learning algorithms, we develop composition–property models for twenty-five properties, which are in agreement with experimental observations. These models are further interpreted using a game-theoretic concept namely, Shapley additive explanations, to understand the role of glass components in controlling the final property. The analysis reveals that the components present in the glass, such as network formers, modifiers, and the intermediates, play distinct roles in governing each of the optical, physical, electrical, and mechanical properties of glasses. Additionally, these components exhibit interdependence, the magnitude of which is different for different properties. While the physical origins of some of these interdependencies could be attributed to known phenomena such as "boron anomaly", "mixed modifier effect", and the "Loewenstein rule", the majority of the remaining ones requires further experimental and computational analysis of the glass structure. Thus, our work paves the way for decoding the "glass genome", which can provide the recipe for discovery of novel glasses, while also shedding light into the fundamental factors governing the composition–structure–property relationships.

**Introduction**

Glasses have been used by humans extensively for last 4500 years, since their discovery in the Mesopotamian and Egyptian civilizations[1–4]. Today, glasses are extensively used for a wide range of applications including touchscreen panels, windshields, bioactive implants[2,3,5], optical fibers and sensors, kitchen and laboratory glassware, thermal insulators, nuclear waste immobilization[6,7], optical lenses, and solid electrolytes[2]. Such widespread applications of glasses are possible due to their unique optical, mechanical, physical, and electrical properties, which can be attributed to the intrinsic disordered structure—a consequence of the melt-quench process through which glasses are prepared. This disordered structure also allows accommodating almost any element in the periodic table, either as an element itself or as a compound such as oxides or halides, in their structure without any stoichiometric constraints[8]. Despite the widespread usage, and flexibility in terms of composition for inorganic glasses, the composition–property relationships in these glasses are poorly understood[8–11]. This poses a major challenge to: (i) understand the physics governing the glass structure and properties, and (ii) develop novel glasses for targeted applications in an accelerated fashion[12].

To address these challenges, recently, machine learning (ML) techniques have been widely used[12–16]. Several ML algorithms have been used in conjunction with available experimental databases to predict properties such as Young's modulus[14,15], liquidus temperature[17], solubility[18], glass transition temperature[16,19], viscosity[20], and dissolution kinetics[13,21]. The largest models presented thus far, predicts, nine properties, with up to 37 input components[22]. However, all these studies have been restricted to oxide glasses. More importantly, these works attempt to use the ML as black-box models providing little insights into the inherent role played by the compositions in controlling the glass properties. In addition, it is well known that the glass components themselves cannot be considered independent. This is exemplified by the well-known "boron anomaly", which can be attributed to the differential coordination numbers of three of four taken by boron depending on the presence of charge compensating sodium atoms. However, the traditional ML models are agnostic to such inherent interdependencies and does not provide any information on the correlation of the input components for a given property[23].

Here, we address these open challenges by employing explainable machine learning algorithms in conjunction with data-driven models. To this extent, relying on a large database of more than 276,000 glass compositions made of 221 components covering 82 elements of the periodic table, we develop machine learning models for 25 glass properties that exhibit excellent agreement with experimental properties. These properties include Abbe number, annealing point, bulk modulus, compressive strength, crystallization temperature, density, dielectric constant, electrical conductivity, Vickers hardness, refractive index, shear modulus, softening point, strain point, liquidus temperature, Littleton temperature, thermal expansion coefficient, tensile strength, glass transition temperature, thermal conductivity, thermal shock resistance, Infrared transmittance, solar transmittance, visible transmittance, ultraviolet transmittance, and Young's modulus. Further, employing Shapley additive explanations (SHAP), we determine the contribution of each of the input components in a glass toward the final property value. The SHAP value provides insights into the role played by each of the components in a glass towards increasing or decreasing the property value from the overall mean. Finally, using the SHAP interaction value, we analyze the interdependencies between each of the input components for a given glass property. We observe that these interdependencies are not only a function of the glass composition, but the property as well.

**Methods**

**Dataset Preparation**

The raw dataset in this work consists of a large experimental dataset compiled from literature including journal papers and patents, SciGlass[24] (an open-source database), and INTERGLAD v8[25]The raw dataset in this work consists of a large experimental dataset compiled from literature including journal papers and patents, SciGlass[24] (an open-source database), and INTERGLAD v8[25] (a commercial database). These glass datasets contain large experimental data (more than 350,000 data entries) from literature with information such as author, journal and experimental conditions. Similar to the traditional practice, we consider the glass composition in terms of the mol% of its components, which are binary compounds such as oxides, halides, and sulfides, of the elements in the periodic table. The final glass composition can have one, two, or multiple components. Here, we consider only oxides and halides as they form the majority of available inorganic glasses. Thus, the dataset for a property comprises of the input components, their respective mol% for each of the glass compositions, and the property value of the given composition. The initial dataset extracted for a particular property (such as density) had duplicate entries, incomplete entries, and inconsistent entries (for example, the sum of glass components not equal to 100%). Such anomalies were removed with appropriate criteria as: (i) inconsistent and incomplete entries were discarded, (ii) duplicate entries were merged with taking the mean of each property corresponding to duplicate entries while avoiding the outliers if present (two times the standard deviation was used as a criterion to remove outliers), and (iii) randomly checking a few entries against their original references. Some glass components were present in very few glass compositions, leading to very sparse input representation. We removed these components and modified the dataset such that each component is present in at least 30 glass compositions. Finally, we split our dataset into 80:20 ration as training, and test set. The test set was used as a holdout set which was used only for the final model evaluation. The training was used for developing ML model using k-fold cross validation. The approach was consistently used to create the dataset for all the 25 properties. The final dataset developed in this fashion for the 25 properties comprised of 221 input components (see SI for details), which covered 82 elements in the periodic table.

**Dataset visualization**

We visualize dataset using the k-means clustering followed by t-SNE embedding. K-means clustering algorithm uses the Euclidean distance in n-dimensions to cluster the glass compositions that are closer together. For example, sodium-silicate glass composition will be closer to calcium-silicate rather than phospho-borate. This leads to the identification of clusters with similar glass components. We employ k-means clustering to group the final dataset glass compositions to separate clusters. It may also lead to glass composition clusters with specific structural, optical, and thermal properties. Further, we use t-SNE[26] to transform input data from n-dimensional space to two-dimensional space. Further, we use t-SNE[26] to transform input data from n-dimensional space to two-dimensional space. Finally, we plot these (as shown in Figure 1(b)) clusters on a 2D plot where each cluster is color-coded according to their cluster number.

**Extreme Gradient Boosted Decision Trees (XGBoost)**

We train the machine learning models using the XGBoost[27] python package. XGBoost stands for extreme gradient boosting of tree-based machine learning models such as tree ensemble[28–30]. It contains tool set for scalable end-to-end tree boosting system, sparsity-based algorithms, and justified weighted quantile sketch for efficient proposal calculation. XGBoost uses K additive function to predict the output as We train the machine learning models using the XGBoost[27] python package. XGBoost stands for extreme gradient boosting of tree-based machine learning models such as tree ensemble[28–30]. It contains tool set for scalable end-to-end tree boosting system, sparsity-based algorithms, and justified weighted quantile sketch for efficient proposal calculation. XGBoost uses K additive function to predict the output as

$$\hat{y}_i = \sum_{k=1}^{K} f_k(x_i), f_k \in F$$

where K in number of regression tree (which uses classification and regression tree (CART[3131])) algorithm) in a tree ensemble, f is function in functional space F, F is set of all possible CARTs and $x_i$ is input feature vector for $i^{th}$ data point in the given dataset $D = \{x_i, y_i\}, i = \{1, 2, 3, \ldots, n\}$ where n is total number of data points. The tree ensemble is created by iteratively adding new regression trees (CARTs) to improve model accuracy. Finding all possible regression trees which improve model accuracy is impractical. Therefore, an optimal regression tree is created from a single node by iteratively adding branches. Adding branches stops if the allowable depth of a tree is reached or the number of samples at a splitting node is equal to one or equal to the minimum number of samples.

**Model development and hyperparameter optimization**

Different ML models were developed for each of the 25 properties. The input features were the glass composition in mol% and the output was the property value in the respective units. For example, for a glass composition $20(Na_2O).80(SiO_2)$, the input feature would be 20 and 80 corresponding to $Na_2O$ and $SiO_2$ and the output would be the property value of interest. We use squared error as our loss function for optimizer. We use two types of boosters namely GBtree and DART. Further, we use Optuna[32] to do hyperparameter optimization for XGBoost tree ensemble models. The range for the hyperparameters is given in the Tables 1, 2, and 3. Finally, the model with the best validation score is selected.Further, we use Optuna[32] to do hyperparameter optimization for XGBoost tree ensemble models. The range for the hyperparameters is given in the Tables 1, 2, and 3. Finally, the model with the best validation score is selected.

| Table 1. XGBoost tree ensemble specific hyperparameters | |
|---|---|
| **Name of hyperparameter** | **Range/List/Value** |
| booster | dart, gbtree |
| lambda | 1e-8 to 1.0 in log scale |
| alpha | 1e-8 to 1.0 in log scale |
| Number of estimators | 300 |
| subsample | 0.7 to 1 |
| colsample_bytree | 0.7 to 1 |
| reg_alpha | 1e-4 to 1 in log scale |
| reg_lambda | 1e-4 to 1 in log scale |

| Table 2. Booster (DART and GBtree) specific hyperparameters | |
|---|---|
| **Name of hyperpameter** | **Range/List/Value** |
| max_depth | 1 to 9 |
| eta | 1e-8 to 1.0 in log scale |
| gamma | 1e-8 to 1.0 in log scale |
| grow_policy | depthwise, lossguide |

| Table 3. DART specific hyperparameters | |
|---|---|
| **Name of hyperpameter** | **Range/List/Value** |
| sample_type | uniform, weighted |
| normalized_type | tree, forest |
| rate_drop | 1e-8 to 1.0 in log scale |
| skip_drop | 1e-8 to 1.0 in log scale |

**Optuna**

Optuna[32] is a hyperparameter optimization software package. It provides a define-by-run programming environment, efficient sampling and pruning algorithms, modular and is easy to scale. Optuna optimizes (minimize or maximize) an objective function which takes a set of hyperparameters (called *trial*) as an input and returns the validation score as output. It uses both relational (namely CMA-ES[33] and GP-BO[34]) and independent (TPE[35]) sampling algorithms to investigate new *trials*. It also uses pruning techniques like Asynchronous Successive Halving[36] (ASHA) for termination of inefficient *trials*. It also provides an interface using which we can provide custom sampling and pruning methods. Finally, we get an optimized set of hyperparameters and trained ML model.Optuna[32] is a hyperparameter optimization software package. It provides a define-by-run programming environment, efficient sampling and pruning algorithms, modular and is easy to scale. Optuna optimizes (minimize or maximize) an objective function which takes a set of hyperparameters (called *trial*) as an input and returns the validation score as output. It uses both relational (namely CMA-ES[33] and GP-BO[34]) and independent (TPE[35]) sampling algorithms to investigate new *trials*. It also uses pruning techniques like Asynchronous Successive Halving[36] (ASHA) for termination of inefficient *trials*. It also provides an interface using which we can provide custom sampling and pruning methods. Finally, we get an optimized set of hyperparameters and trained ML model.

**Shapley additive explanations for tree ensemble**

Shapley additive explanation[37] (SHAP) values is a unified game theoretic approach to calculate the feature importance of an ML model. SHAP measures a feature's importance by quantifying the prediction error while perturbing a given feature value. If the prediction error is large, the feature is important, otherwise the feature is less important. It is an additive feature importance method which produces unique solution while adhering to desirable properties namely local accuracy, missingness, and consistency. SHAP introduces model agnostic approximation methods such as Kernel SHAP as well as model-specific approximation methods such as Deep SHAP, Max SHAP, and Linear SHAP, etc. Here, we use a tree-specific SHAP approximation method namely TreeExplainer[38]. Shapley value calculation requires a summation over all possible feature subsets, which leads to an exponential time complexity. TreeExplainer exploits the internal structure of tree-based models and collapse summation to a set of calculation specific to the leaf node of a tree model leading to low order polynomial time complexity[38]. TreeExplainer also calculates interaction values to capture the local interaction effects. These local interaction values can be used to explain the effects of feature interaction[39]. Shapley additive explanation[37] (SHAP) values is a unified game theoretic approach to calculate the feature importance of an ML model. SHAP measures a feature's importance by quantifying the prediction error while perturbing a given feature value. If the prediction error is large, the feature is important, otherwise the feature is less important. It is an additive feature importance method which produces unique solution while adhering to desirable properties namely local accuracy, missingness, and consistency. SHAP introduces model agnostic approximation methods such as Kernel SHAP as well as model-specific approximation methods such as Deep SHAP, Max SHAP, and Linear SHAP, etc. Here, we use a tree-specific SHAP approximation method namely TreeExplainer[38]. Shapley value calculation requires a summation over all possible feature subsets, which leads to an exponential time complexity. TreeExplainer exploits the internal structure of tree-based models and collapse summation to a set of calculation specific to the leaf node of a tree model leading to low order polynomial time complexity[38]. TreeExplainer also calculates interaction values to capture the local interaction effects. These local interaction values can be used to explain the effects of feature interaction[39].

The SHAP values are visualized using a violin plot and a river flow plot. The violin plot represents the contribution of a given feature towards the different output values as a function of the feature value. Thus, the violin plot is colored according to the feature value. The river plot shows some specific paths for the final prediction, where the intermediate points represent the contribution of different input components. These paths are created by nudging the prediction from expected value towards a particular direction representing that specific glass component's particular contribution. Thus, the river flow plot is colored according to the final output value.

**Results and discussion**

Inorganic glasses are traditionally made by the melt-quenching of a mixture of one, two, or multi-component binary compounds such as oxides, halides, or sulfides, for example, $SiO_2$ or NaF. To develop machine learning (ML) models for the composition–property relationships, we compiled a large experimental dataset comprising of 275,516 glass compositions made of 221 distinct components. Figure 1(a) shows the 82 elements present in our dataset in the form of binary compounds along with the roles of cations in the glass network, namely, network former (NF), modifier (NM), or intermediate (NI). Since all the properties are not available for every composition, the dataset is different for each property. To visualize, the complete input feature space of our dataset, we combine all the glass compositions that are present in our dataset for which at least one property is available.

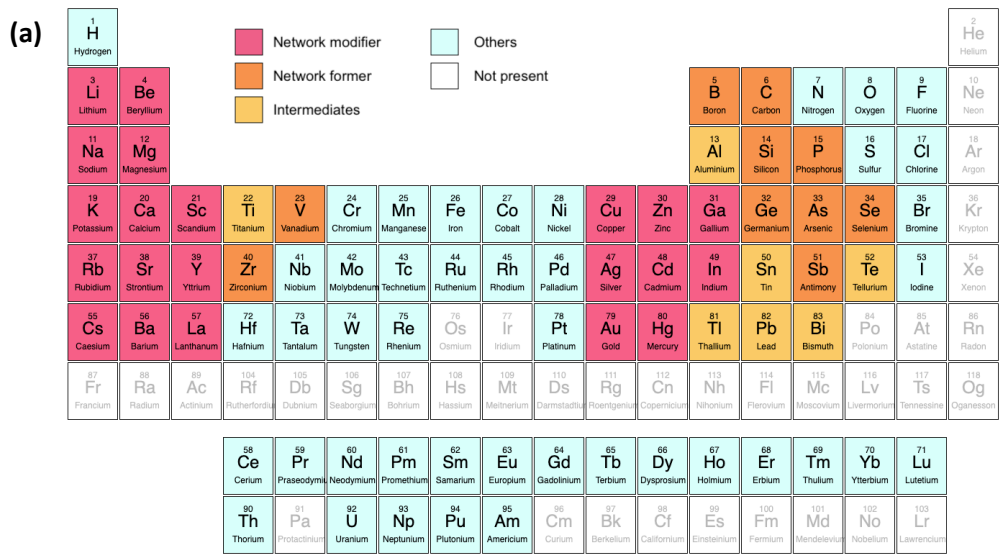

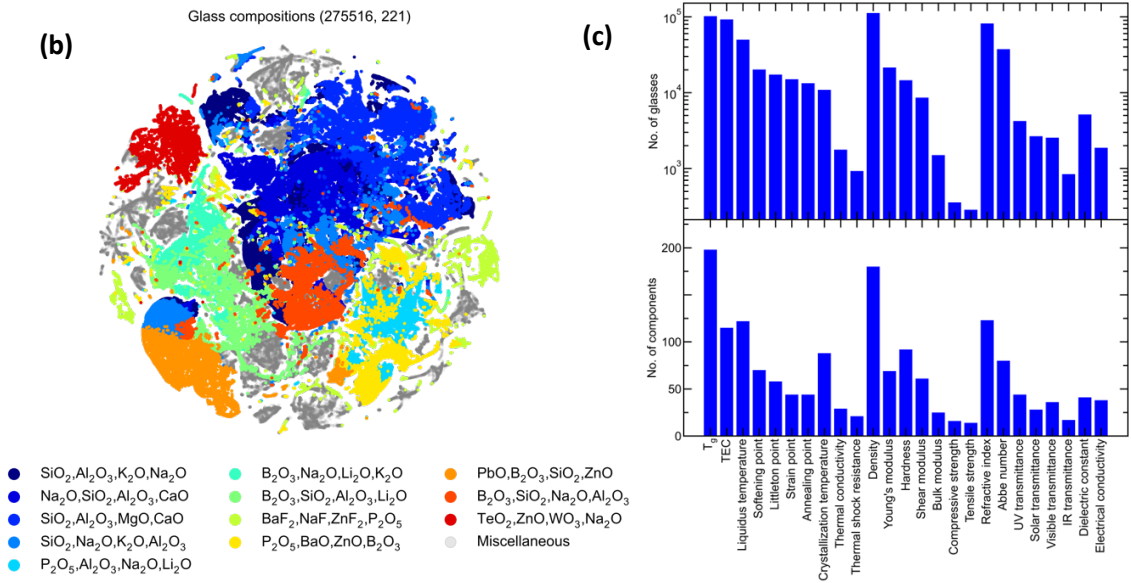

**Figure 1. Dataset visualization.** (a) Periodic table with the elements present in our dataset highlighted as network formers, modifiers, intermediate, and others. (b) Two-dimensional t-SNE embedding of all the glass compositions used in the present work comprising a total of 275,516 compositions and 221 components. The major families are highlighted as per the scheme shown in the legend. (c) Bar plot of the total number of glass compositions (upper subplot) and components (lower subplot) for the properties present in our dataset.

Figure 1(b) shows the two-dimensional t-SNE plot of all the glass compositions, color-coded based on the clusters obtained through k-means clustering. The top 12 largest clusters are labeled with the four most abundant glass components in the decreasing order of their presence, while the remaining clusters are combined together in a single cluster as miscellaneous. As expected, we note that the glass compositions with similar components are found to be clustered together. All the glass compositions with $SiO_2$ as the major component are grouped as neighbors with significant overlap, while borosilicates are grouped near the silica family. Some of the clusters could also be identified based on their properties as well. For example, fluoride glasses, known for their crystallization behavior, form an isolated cluster. This observation is further confirmed by the t-SNE embedding plotted for each property separately (see SI).

Figure 1(c) (upper subplot) shows the bar plot of the number of glass compositions for each property. For most of the properties the number is more than 10,000, while for a few properties such as thermal shock resistance, compressive strength, and tensile strength, the number is less than 1000. Figure 1(c) (lower subplot) shows the number of glass components (input features) associated with each property. This number represents the union of all the components present in the glasses for a given property; the individual glass compositions themselves could be one-, two-, or multi-component. Here, properties like density and glass transition temperature show a higher number of glass components as these measurements are done for most of the glasses. Whereas properties such as transmittance have a lower number of glass components, these are only measured for optical glasses. As expected, the number of components is proportional to the number of glass compositions for most of the properties.

**Composition–property models**

Figure 2 shows the measured values for the 25 properties with respect to the values predicted by the extreme gradient boosted decision tree (XGBoost) models (see Methods for details). Due to the large number of overlapping points, we plot a heat map, where the color of each pixel represents the number of data points in that region. Only the test data is shown in the figure for a fair representation of the model performance. The training and test $R^2$ scores are also provided for each property. As expected, most of the properties exhibit excellent agreement with experiment with high $R^2$ values on the test set (> 0.80). These include Abbe number, annealing point, crystallization temperature ($T_c$), density, dielectric constant, electrical conductivity, Vickers harness, shear modulus, softening point, liquidus temperature ($T_m$), thermal expansion coefficient (TEC), tensile strength, glass transition temperature ($T_g$), infrared (IR) transmittance, solar transmittance, visible transmittance, and ultraviolet (UV) transmittance. However, some of the properties, namely bulk modulus, thermal shock resistance and thermal conductivity, show a wider spread with test set $R^2$ values ranging between 0.70 and 0.80, suggesting a relatively poor performance of the XGBoost models. This could be attributed to the following. (1) Bulk modulus is mostly determined indirectly for glasses from the other elastic properties or measurements, which adds increased noise in the dataset. (2) Thermal shock resistance is the maximum temperature difference at which glass exhibits fracture, which is highly sensitive to the least count in the temperature measurements. Overall, we demonstrate that the ML models are able to capture the composition–property relationships in a reasonable fashion.

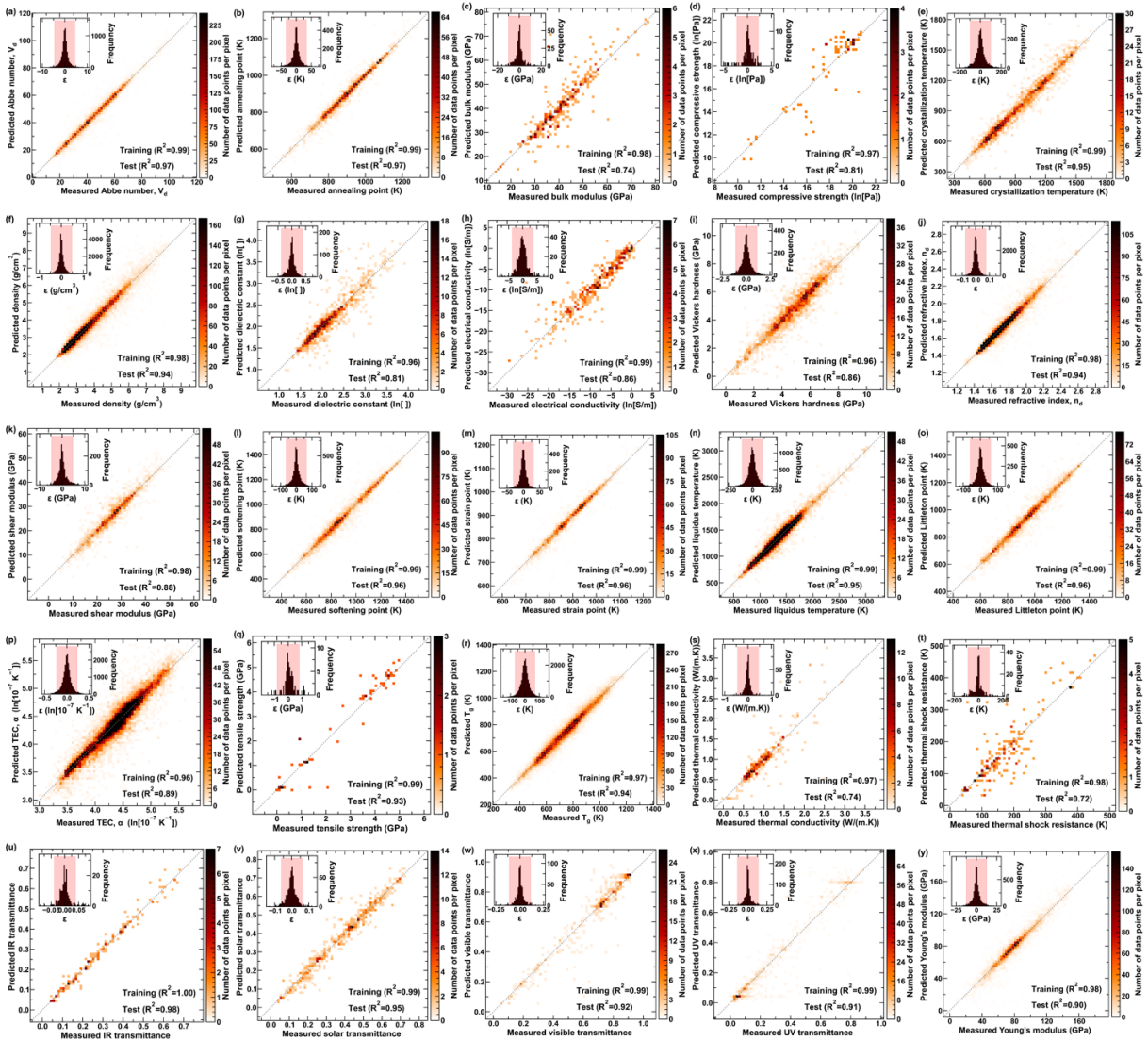

**Figure 2. Composition–property models.** Predicted values with respect to measured values for (a) Abbe number, (b) annealing point, (c) bulk modulus, (d) compressive strength, (e) crystallization temperature, (f) density, (g) dielectric constant, (h) electrical conductivity, (i) Vickers hardness, (j) refractive index, (k) shear modulus, (l) softening point, (m) strain point, (n) liquidus temperature, (o) Littleton temperature, (p) thermal expansion coefficient (TEC), (q) tensile strength, (r) glass transition temperature, (s) thermal conductivity, (t) thermal shock resistance, (u) infrared (IR) transmittance, (v) solar transmittance, (w) visible transmittance, (x) ultraviolet (UV) transmittance, and (y) Young's modulus for test dataset. Heat map shows number of datapoints per pixel according to respective color scheme. The inset shows the histogram of error and shaded region represents the 95% confidence interval. Test and training $R^2$ values are also provided.

### Interpreting the composition–property models

Now, we use the Shapley additive explanations (SHAP), a game-theoretic post-hoc model-agnostic approach, to interpret the compositional control of the properties. For each composition, SHAP quantifies the contribution of each glass component to the final prediction with respect to mean value of the property. This allows both qualitative and quantitative interpretation of the role played the glass components in controlling a given property. Figure 3 shows the SHAP values for top 20 glass components for nine properties, namely, $T_g$, thermal expansion coefficient, liquidus temperature, density, Young's modulus, hardness, refractive index, shear modulus, and dielectric constant (see SI for all 25 properties). The components are arranged with the mean absolute SHAP values increasing from left to right. Thus, these components represent the ones that significantly impact the property value, either in increasing or decreasing the model output from the mean value.

Figure 3 shows the SHAP values of the input components and how it leads to the final output value through the violin (lower subplot) and the river-flow plot (upper subplot), respectively. For the violin scatter plot, the thickness and color represent the density of points the input feature value, respectively. For the river-flow plot, the color and the path of the line represent the output value and the contribution of individual input components, respectively. Thus, the violin plot focusses on the contribution of a given input feature toward different property values, whereas the river-flow plot focusses on the contributions of different input features toward a given property value. Hence, this approach allows model interpretation from the perspective of both composition and property.

We can divide the glass components into four categories: (C1) SHAP value increases with feature value, (C2) SHAP value decreases with feature value, (C3) mixed effects, and (C4) SHAP values are either always positive or negative. Care should be taken while comparing SHAP values as the contribution of an input to the prediction is quantified (positive or negative) with respect to the mean output. For example, in Figure 3(a), the SHAP value of CaO is always positive implying that the glass transition temperature is higher with respect to the expected model output with higher CaO content. However, traditionally in glass science, such comparison is usually done with respect to pure silica, where the glass transition temperature decreases with the addition of CaO, or with respect to any other base glass. Thus, the SHAP values in Fig. 3 should be read in conjunction with the mean property value to understand the effect of a component.

We observe that for each property, there may be components belonging to C1, C2, C3 and C4. Interestingly, we observe that the top five components governing the output always consists of at least one among the NFs, NMs and NIs for almost all the properties, with $SiO_2$ being almost always present among them (see Fig. 3, and SI). The impact of these components on the properties are distinct and depends on the underlying physics governing the structure–property correlation. To illustrate this, we take the specific example of $T_g$.

Figure 3(a) shows that glass components, namely $Al_2O_3$, $La_2O_3$, CaO, $Nb_2O_5$, $TiO_2$, $GeO_2$, BaO, MgO, $ZrO_2$, and $Ga_2O_3$, have positive SHAP values and belong to C4. Thus, higher content of these components in the glass composition will produce glasses with higher $T_g$ with respect to the expected model output. In contrast, when present in lower quantity, they do not affect the output significantly. Glass components, namely $Na_2O$, $Li_2O$, PbO, $K_2O$, $Ag_2O$, $Bi_2O_3$, and $V_2O_5$, have negative SHAP values and belong to C4. Higher content of these components in glass composition will produce glasses with lower $T_g$ than the expected model output. In contrast, they do not affect the output significantly when present in lower quantity. In the case of $SiO_2$, the SHAP value increases with increasing SiO$_2$ content (belongs to C1). Glass components P$_2$O$_5$ and B$_2$O$_3$ show mixed effects and belongs to C3. Mixed effects in components suggest that the behavior of the component may also be governed by the other components present in the glass. Indeed, "boron anomaly" is well-known in glass science, where the glass transition (and viscosity) changes non-monotonically with the B$_2$O$_3$ content as boron can be present in two coordination states namely B$^{3+}$ and B$^{4+}$, depending on the absence or presence, respectively, of charge compensating cations in the vicinity. The case of P$_2$O$_5$, although may require further investigation, could be attributed to the differential interactions with the cations due to the P–O bond length variations in the orthophosphate species.

Similar compositional control is observed in the case of $T_m$ and $T_c$ as well with SiO$_2$ belonging to C1 and NMs or NIs such as Al$_2$O$_3$, La$_2$O$_3$, CaO, Nb$_2$O$_5$, TiO$_2$, BaO, MgO, and ZrO$_2$ belonging to C4 with positive SHAP values. NMs such as Na$_2$O, Li$_2$O, PbO exhibit negative SHAP values for $T_c$. However, interestingly, NMs and NFs such as Na$_2$O, Li$_2$O, PbO, K$_2$O, and B$_2$O$_3$ exhibit mixed behavior suggesting correlated effects of other input components. A similar compositional effect can also be found on other physical properties such as annealing, strain, softening, and Littleton point, as they are all related to the structural relaxation of glasses with respect to the temperature.

In Fig. 3(b), NFs and NIs such as SiO$_2$, Al$_2$O$_3$, and GeO$_2$ belong to C2 implying lower TEC values for glass compositions with these components. NMs such as Na$_2$O, K$_2$O, CaO, Li$_2$O, BaO belong to C4 implying higher TEC with increasing values of these components in the glass composition. This observation is in agreement with the literature in glass science, which suggests that the glass modifier tend to open up the structure leading to higher TEC value for glasses[40–42].

In mechanical properties, we observe that the density is governed by heavier elements and NFs. While NFs decrease the density, heavier elements increase the density of the glass. Further, NFs or NIs such as SiO$_2$, and Al$_2$O$_3$, enhance the hardness, while alkali modifiers such as Na$_2$O, and K$_2$O reduce the hardness. For bulk modulus, most NMs enhance the value due to better packing, while the NFs have mixed effects. For optical properties, the components increasing the value the refractive index tend to decrease the Abbe number and vice-versa. Thus, the detailed compositional control on the inorganic glass properties is revealed through the analysis. While the role of some of these components are known for some properties in the literature, majority of the information revealed through the SHAP analysis (see Fig. 3) is hitherto unknown and remain to be explored. Although the validity of the results is established as the model relies on the experimental data, understanding the mechanism related to many of these composition–property relationships require further exploration.

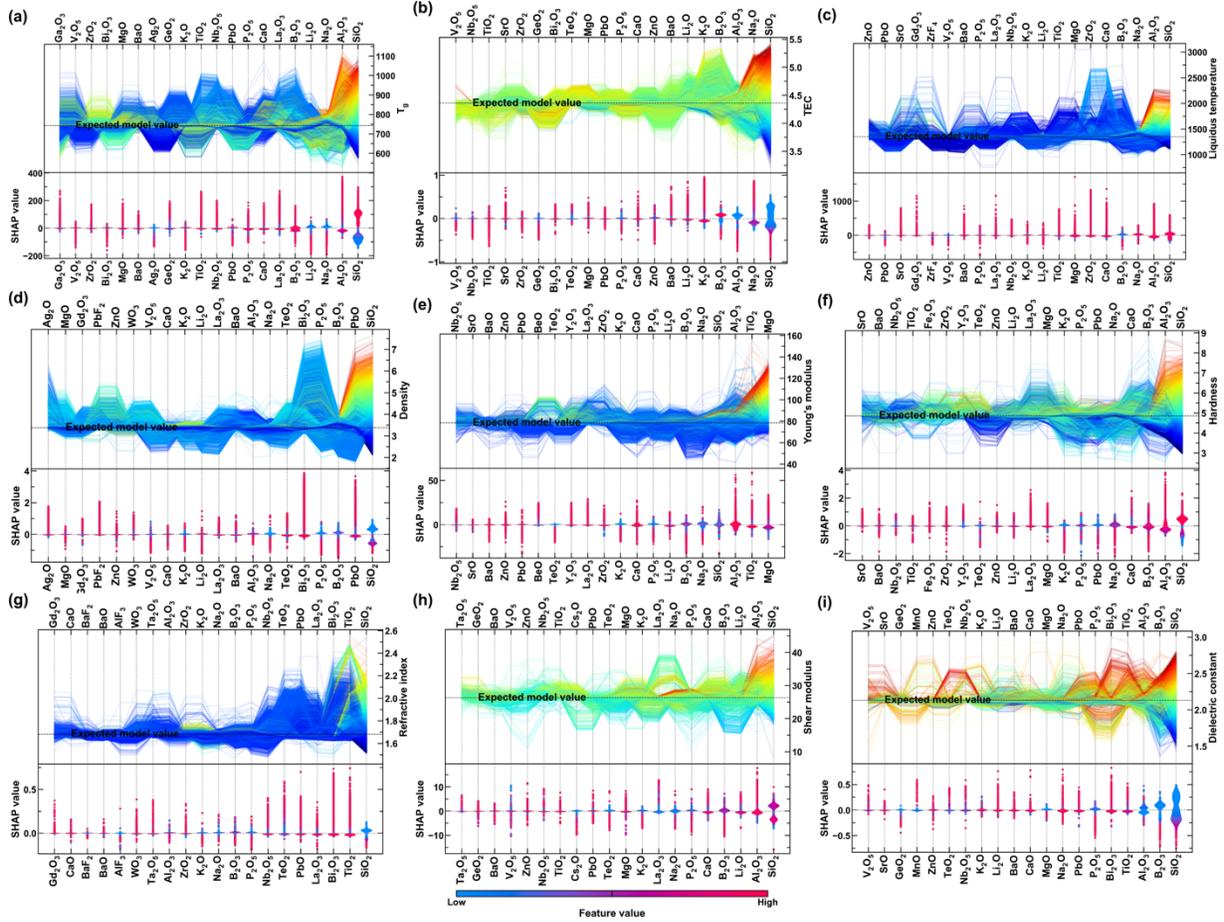

**Figure 3. Interpreting the composition–property models.** SHAP violin and river flow plots for (a) glass transition temperature, (b) thermal expansion coefficient, (c) liquidus temperature, (d) density, (e) Young's modulus, (f) hardness, (g) refractive index, (h) shear modulus and (i) dielectric constant (see SI for all 25 properties)

**Decoding the glass component correlations**

While the composition–property relationships can still be explored using parametric studies, a lesser understood aspect in glass science is the correlation between the components present in a glass composition. Understanding these correlations require very high precision experiments such magic angle spinning nuclear magnetic resonance studies, or advanced computational simulations such as density functional theory (DFT) modeling of glass structure. Here, using the SHAP interaction values, we identify the correlation between each of the input components in governing the output value. To this extent, we analyze the error in the output prediction while perturbing two input components simultaneously. If the magnitude of error in the output while perturbing a single input component is the same for different values of the second output components, this suggests that the two input components are not correlated; otherwise, they are correlated. The degree of correlation can also be quantified from the SHAP interaction values (see Methods).

Figure 4 shows the interaction values for top (20) glass components for $T_g$, TEC, $T_m$, density, Young's modulus, hardness, refractive index, shear modulus and dielectric constant (see SI for all 25 properties). The glass components are arranged in the decreasing value of the mean absolute SHAP from left to right, while the color shows the normalized interaction values for glass components. Similar to the SHAP values, we observe that the correlation between the different input components is different for each property. This suggests that while some effects such as "boron anomaly" may have a significant effect on the properties such as hardness, glass transition temperature, liquidus temperature (as exemplified by the high interaction values, see Fig. 4), it may not have a notable effect of properties such as refractive index or dielectric constant (as exemplified by low interaction values).

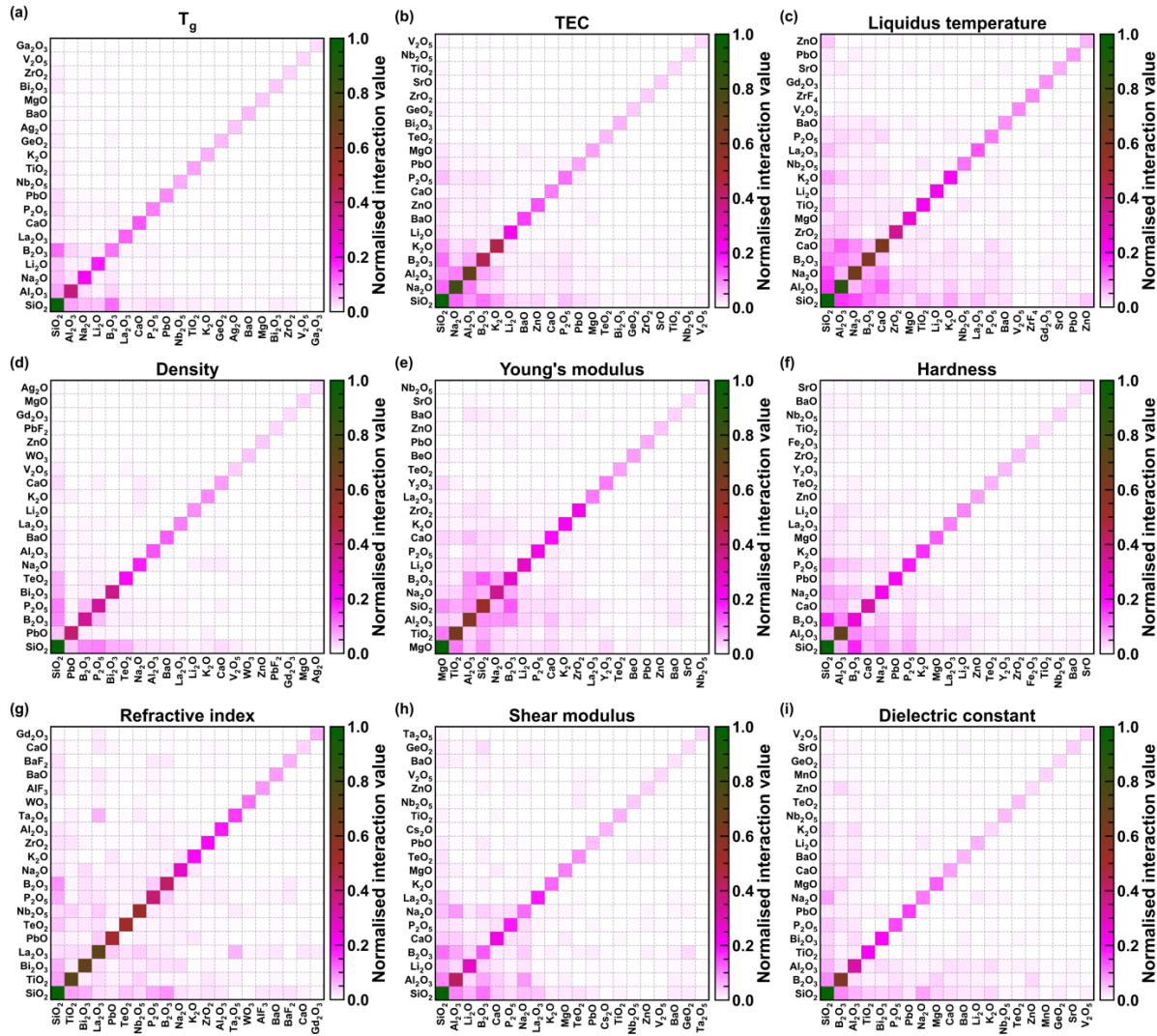

**Figure 4. Interpreting the correlation of input features.** Heat map of normalized SHAP interaction values for top 20 glass components for (a) glass transition temperature, (b) thermal expansion coefficient, (c) liquidus temperature, (d) density, (e) Young's modulus, (f) hardness, (g) refractive index, (h) shear number and (i) dielectric constant (see SI for all 25 properties).

Further, the three well-known compositional correlations in glass science, namely, boron anomaly, mixed-modifier (MM) effect, and Loewenstein's rule are clearly captured by the SHAP interaction values. The MM effect is confirmed by the high interaction values between alkali and alkaline earth species for Young's modulus, hardness, TEC, $T_m$, and bulk modulus, to name a few. Similarly, the Loewenstein's rule is confirmed by the high interaction values between alumina and silica, suggesting that the structure and properties of aluminosilicate glasses are controlled by the interaction between silica and alumina. In addition to these known interrelationships, Fig. 4 reveals several other correlations between the glass components, the origin of which requires detailed experimental and computational modeling.

Altogether, relying on a large dataset, the present work develops the largest composition–property models of the inorganic glasses. The detailed analysis reveals that interpretable machine learning can be used to explain and understand the hidden composition–property relationship in inorganic glasses. In addition, the correlation between several input features explained using the interpretable machine learning techniques can provide guidance in gaining further insights into the physics governing composition–structure–property relationships. However, the detailed understanding of the specific mechanisms requires further high-fidelity experiments or simulations. Overall, the present work provides deep insights into the "genome" of glasses, thereby providing a recipe for accelerated development of glasses as well as guidance for understanding the physics of glass properties.

**Acknowledgement**

NMAK acknowledges the financial support for this research provided by the Department of Science and Technology, India, under the INSPIRE faculty scheme (DST/INSPIRE/04/2016/002774) and DST SERB Early Career Award (ECR/2018/002228). RR acknowledges the financial support from Prime Minister's Research Fellowship. The authors thank the IIT Delhi HPC facility for providing the computational and storage resources.

**Code availability**

All the codes used in the present work are available at: https://github.com/m3rg-repo/machine_learning_glass/tree/master/Revealing_the_Compositional_Control_of_Inorganic_Glass_Properties

**Data availability**

The data that support the findings of this study are available from the corresponding author upon reasonable request.